\newcommand{\half}{\mbox{$\frac{1}{2}$}}
\documentclass[aps,prl,groupedaddress,showpacs,nofootinbib,
              preprint,showpacs,showkeys]{revtex4}
\usepackage{graphicx}
\usepackage{amsmath}
\usepackage{bm}

\begin{document}
\preprint{ANL-HEP-PR-02-035}

\title{ Radiative decay of $\bm{\Upsilon(nS)}$ into $\bm{S}$-wave
sbottomonium}


\author{Edmond L. Berger, Geoffrey T. Bodwin, and Jungil Lee}
\affiliation{
HEP Division,
Argonne National Laboratory, 9700 South Cass Avenue, Argonne, Illinois
60439 }


\date{\today}

\begin{abstract}
A calculation is presented of the radiative decay of the $\Upsilon(nS)$
into a bound state of bottom squarks.  Predictions are provided of the
branching fraction as a function of the masses of the bottom squark and
the gluino.  Branching fractions as large as several times $10^{-4}$ are
obtained for supersymmetric particle masses in the range suggested by
the analysis of bottom-quark production cross sections.  Data
are shown that limit the range of allowed masses.  Forthcoming 
high-statistics data from the CLEO Collaboration offer possibilities  
of discovery or significant new bounds on the existence and masses of
supersymmetric particles.
\end{abstract}


\maketitle


A long-standing puzzle in high-energy strong interactions is the fact
that the rate for bottom quark ($b\bar{b}$) production at the Fermilab
Tevatron collider is two to three times greater than the theoretical
prediction from quantum chromodynamics (QCD)
\cite{Frixione:1994nb}.  The next-to-leading order QCD contributions are 
large, and a combination of further higher-order effects in production 
and/or fragmentation may eventually reduce the discrepancy~\cite{frag}.
An alternative explanation is offered in
Ref.~\cite{Berger:2000mp} in the context of physics beyond the standard
model. There, it is argued that a solution may be provided by the
existence of a light bottom squark $\tilde b$ and a light gluino
$\tilde{g}$, with masses in the ranges $2~\hbox{GeV} < m_{\tilde{b}}
<5.5~\hbox{GeV}$ and $12~\hbox{GeV} < m_{\tilde{g}} <16~\hbox{GeV}$. The
$\tilde g$ and the $\tilde b$ are the spin-1/2 and spin-0 supersymmetric
partners of the gluon ($g$) and bottom quark ($b$). The masses of all
other supersymmetric particles are assumed to be arbitrarily heavy,
i.e., of order the electroweak scale or greater~\cite{Berger:2000mp}.
While this scenario is not among the more popular schemes for
supersymmetry breaking, the hypothesis of a light bottom squark is not
inconsistent with direct experimental searches and indirect constraints
from other
observables~\cite{light-sb,CLEO-sb,DELPHI-sb,Berger,Berger:2002vs}. 
Therefore, it is essential either to confirm
or to refute this proposal by examining its implications for additional
processes.  In this Letter, we show that high-statistics data being
accumulated and analyzed now at the Cornell Electron Storage Ring (CESR)
facility~\cite{CLEO-c} could provide definitive confirmation of the
proposal of Ref.~\cite{Berger:2000mp} or severely constrain the allowed
parameter space~\cite{directdecays}.

In the proposal of Ref.~\cite{Berger:2000mp}, it is possible that the bottom 
squark is relatively stable and, hence, bound states of a bottom 
squark and bottom antisquark (sbottomonium) could exist. These bound states 
could be produced in radiative decays of bottomonium states, such as 
$\Upsilon \rightarrow \tilde{S} \gamma$, where $\tilde{S}$ is the $S$-wave 
bound state of a $\tilde{b} \tilde{b}^*$ pair.  Alternatively, the $\tilde{b}$ 
could decay promptly via $R$-parity and baryon-number 
violation~\cite{Berger:2000mp,Berger}, and no bound state would be formed.   

In this Letter, we compute the rate for an $\Upsilon(nS)$ to decay
radiatively into an $S$-wave $\tilde{b} \tilde{b}^*$ bound state. 
We show that, provided that a bound state is formed, the resonance search 
by the CUSB Collaboration~\cite{CUSB-gamma} already increases the allowed 
lower bounds on $m_{\tilde{b}}$ and $m_{\tilde{g}}$.  
Discovery of the bound states may be possible with the high-statistics 
2002 CLEO-c data set, or a larger 
range of bottom-squark and gluino masses may be disfavored.

The mass eigenstates of the bottom squarks, $\tilde{b}_1$ and 
$\tilde{b}_2$, are mixtures of $\tilde{b}_L$ and $\tilde{b}_R$, the 
supersymmetry partners the left-handed (L) and right-handed (R) bottom
quarks:
\begin{subequations}
\begin{eqnarray}
|\tilde{b}_1\rangle = \sin\theta_{\tilde{b}}|\tilde{b}_L\rangle +
                                \cos\theta_{\tilde{b}}|\tilde{b}_R\rangle, \\
|\tilde{b}_2\rangle = \cos\theta_{\tilde{b}}|\tilde{b}_L\rangle -
                                \sin\theta_{\tilde{b}}|\tilde{b}_R\rangle.
\label{eq:mixing}
\end{eqnarray}
\end{subequations}
We take $\tilde{b}_1$ to be the eigenstate of lighter mass, and we drop
the subscript 1 in the remainder of this paper. The mixing angle
$\theta_{\tilde{b}}$ is constrained by the requirement that the coupling
of a $\tilde b\tilde b^*$ pair to the $Z$ boson be sufficiently small to
be compatible with data~\cite{light-sb}. At lowest order (tree-level), this 
requirement implies that 
$\sin^2\theta_{\tilde{b}} \approx 1/6$~\cite{Cao:2001rz}.
\footnote{
In the first paper of Ref.~\cite{Cao:2001rz}, it is argued that one-loop
contributions may render the light $\tilde{g}$ and light $\tilde{b}$
scenario inconsistent with data at the $2\sigma$ level, unless the mass
of the heavier ${\tilde{b}}_2$ is less than about 125 GeV.  Making
somwhat different assumptions, the authors of the second paper obtain a
$5\sigma$ bound of 180 GeV and the author of the third paper obtains a
$3\sigma$ bound of more than 200 GeV. The possibility that the mass of
the ${\tilde{b}}_2$ could be as low as 100 GeV or so is not excluded by
data.  Analysis of data from $e^+ e^-$ interactions at LEP, for example,
would require first a detailed modeling of the decay modes of the
${\tilde{b}}_2$.}

\begin{figure}
\includegraphics[width=8cm]{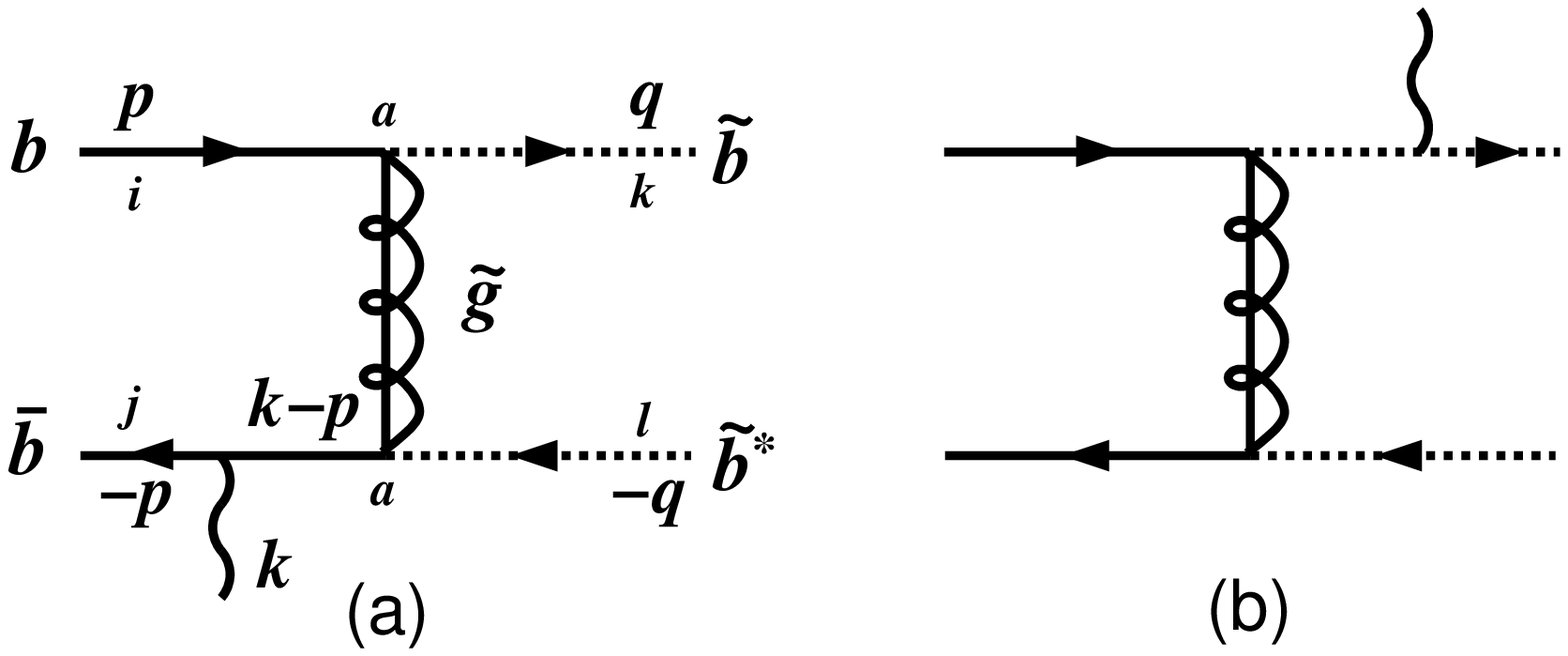}
\caption{Feynman diagrams for the process
$b\overline{b}\to \tilde{b} {\tilde{b}}^*\gamma$. }
\label{fig:feynman}
\end{figure}

We calculate the decay rate $\Gamma(\Upsilon\rightarrow
\tilde{S} \gamma)$ in the framework of the nonrelativistic QCD (NRQCD)
factorization formalism \cite{BBL}.
\footnote{One can adapt the NRQCD
formalism for spin-1/2 quarks to the case of spin-0 squarks by dropping
the spin-dependent interactions and replacing Pauli fields with scalar
fields.}
First, we
compute the amplitude in full QCD for the process $b(p)+\bar{b}(p)\to
\tilde{b}(q) +\tilde{b}^{*}(q)+\gamma(k)$. Typical Feynman diagrams 
are shown in Fig.~\ref{fig:feynman}. The indices $i$, $j$,
$k$, and $l$ label the colors of the incident $b$ and $\bar{b}$ and the
final $\tilde{b}$ and $\tilde{b}^*$, respectively. The color index of the
exchanged gluino is $a$.  The Feynman rules~\cite{directdecays} depend on the 
mixing angle $\theta_{\tilde{b}}$.

We carry out the computation in the $b\bar b$ rest frame and choose the
radiation gauge. Then, the photon-$\tilde{b}$-$\tilde{b}^*$ vertex
vanishes, since it is proportional to $\epsilon^*\cdot
(2q+k)=\epsilon^*\cdot (2p)$, where $\epsilon$ is the polarization of
the photon. Therefore, we need to compute only the two diagrams of the
type in Fig.~\ref{fig:feynman}(a), in which the photon attaches to
the $b$ or the $\bar{b}$. We make use of the spin-triplet projector
\cite{projop}
\begin{eqnarray}
\sum_{\lambda_1,\lambda_2}u({\bf p},\lambda_1)\bar{v}(-{\bf p},\lambda_2)
\langle \half\,\lambda_1\,\half\, \lambda_2
|1\,\epsilon_\Upsilon\rangle
=-\frac{1}{\sqrt{2}}\not\!\epsilon_\Upsilon(\not\! p-m_b),
\end{eqnarray}
where $\epsilon_\Upsilon$ is the polarization of the $b\bar b$ system,
the Dirac spinors $u(p,s_1)$ and $v(p,s_2)$ have the relativistic
normalization $\bar u(p,s) u(p,s)=\bar v(p,s) v(p,s) = 2E_b(p)$, and
$E_i(p)=\sqrt{{\bf p}^2_i+m_i^2}$.
Projecting onto color-singlet $b\bar b$ and $\tilde b \tilde b^*$ states, 
which leads to the color factor
$(\delta_{ij}\delta_{kl}/N_c)T^a_{ki}T^a_{jl}=\frac{4}{3}$,
we find that the amplitude in full QCD is 
\begin{eqnarray}
\mathcal{M}=i\frac{16\sqrt{2}}{3m_{\tilde{g}}} g_s^2 e e_b 
           \sin\theta_{\tilde{b}}\cos\theta_{\tilde{b}}\;
           \epsilon\cdot\epsilon_\Upsilon .
\end{eqnarray}
We use plane-wave states for the quarks, squarks, and
photon, without any factors $1/(2E)$. Squaring the matrix element
and averaging over the polarizations of the $\Upsilon$, we obtain
\begin{eqnarray}
\overline{|\mathcal{M}|^2}=
\frac{1024}{27 m_{\tilde{g}}^2} g_s^4 e^2 e^2_b 
           \sin^2\theta_{\tilde{b}}\cos^2\theta_{\tilde{b}}.
\label{me-squared}
\end{eqnarray}

We match the squared matrix element in Eq.~(\ref{me-squared}) onto NRQCD 
for the $b\bar{b}$ system in the $b\bar b$ rest frame and onto NRQCD for the 
$\tilde{b} \tilde{b}^*$ system in the $\tilde b \tilde b^*$ rest frame 
using, respectively, 
\begin{eqnarray}
\overline{|{\cal M}|^2}&=&
F_1({}^3S_1)\langle b\bar b|{\cal O}_1({}^3S_1)|b\bar b\rangle,\\
\overline{|{\cal M}|^2}&=&
\tilde F_1({}^1S_0)\langle 0|\tilde{\cal O}_1^{\tilde b\tilde b^*}
({}^1S_0)|0\rangle.
\end{eqnarray}
Here, $F_1$ and $\tilde F_1$ are short-distance coefficients, 
${\cal O}_1({}^3S_1)$ is a four-quark operator, and $\tilde{\cal O}_1^{\tilde 
b\tilde b^*}({}^1S_0)$ is a four-squark operator: 
\begin{eqnarray}
\mathcal{O}_1(^3S_1)&=&
\psi^\dagger \sigma^k \chi\chi^\dagger \sigma^k\psi,
\\
\tilde{\mathcal{O}}_1^{\tilde{b}\tilde{b}^*}(^1S_0)&=&
\tilde{\chi}^*\tilde{\psi}
\sum_{X}|\tilde{b}\tilde{b}^*+X\rangle\langle\tilde{b}\tilde{b}^*+X|
\tilde{\psi}^*\tilde{\chi} .
\end{eqnarray}
The Pauli field $\psi$ annihilates a $b$ quark, the Pauli field
$\chi$ creates a $b$ antiquark, the scalar field $\tilde\psi$
annihilates a $b$ squark, the scalar field $\tilde \chi$ creates a $b$
squark, and the sum is over all intermediate states $X$.  The
short-distance coefficient $F_1({}^3S_1)$ for the $b\bar b$ operator
contains $\langle 0|\tilde{\cal O}_1^{\tilde b\tilde
b^*}({}^1S_0)|0\rangle$, and the short-distance coefficient for the
$\tilde b \tilde b^*$ operator $\tilde F_1({}^1S_0)$ contains $\langle
b\bar b|{\cal O}_1({}^3S_1)|b\bar b\rangle$. 
In the Born approximation, the matrix elements are 
\begin{eqnarray}
\langle b\bar b|{\cal O}_1({}^3S_1)|b\bar b\rangle &=&
2(2E_b)^2N_c,\\
\langle 0|\tilde{\cal O}_1^{\tilde b\tilde b^*}
({}^1S_0)|0\rangle &=& (2\tilde{E}_{\tilde{b}})^2 N_c,
\end{eqnarray}
where $\tilde{E}_{\tilde{b}}$ is the energy of
$\tilde{b}(\tilde{b}^*)$ in the $\tilde{b}\tilde{b}^*$ rest frame.
The factors $2E_b$ and $2\tilde{E}_{\tilde{b}}$ appear 
because the free-particle states are normalized  to
$2E$ particles per unit volume.
The factors $N_c$ come from the color traces, and the
additional factor $2$ in the $b\bar b$ case comes from the spin trace.
From this matching process, we deduce that
\begin{eqnarray}
\overline{|\mathcal{M}|^2}
=
\frac{512}{243 m_{\tilde{g}}^2(2E_b)^2(2\tilde{E}_{\tilde{b}})^2} 
g_s^4 e^2 e^2_b
           \sin^2\theta_{\tilde{b}}\cos^2\theta_{\tilde{b}}\;
\langle b\bar{b}| \mathcal{O}_1(^3S_1)
                |b\bar{b}\rangle
\langle 0| \tilde{\mathcal{O}}_1^{\tilde{b}\tilde{b}^*}(^1S_0)
                            |0\rangle.
\end{eqnarray}

To compute the decay rate of the process $\Upsilon\to \tilde{S} \gamma$,
we replace the $b\bar b$ state by the $\Upsilon$ state and replace the
$\tilde b \tilde b^*$ state by the $\tilde S$ state in the squared
matrix element, multiply by the two-body phase
$\Phi_2=(1/8\pi)[1-(M_{\tilde S}^2/M_\Upsilon^2)]$, and multiply by
$2M_{\tilde S}$, where
$M_{\tilde S}$ is the mass of the $\tilde b \tilde b^*$ bound state, and
$M_\Upsilon$ is the $\Upsilon$ mass.  
\footnote{The factor $2M_{\tilde{S}}$ appears for the following reason.
The $\tilde S$ state is normalized to one particle per unit volume in
the $\tilde S$ rest frame. In order to preserve that normalization in
the $\Upsilon$ rest frame, one must multiply by the 
Lorentz-contraction factor for the volume, namely,
$2M_{\tilde{S}}/(2E_{\tilde{S}})$, where 
$E_{\tilde{S}}$ is the energy of the $\tilde{S}$ in the $\Upsilon$ rest frame.
The factor 
$1/(2E_{\tilde{S}})$ is absorbed
into the conventional definition of the phase space.}
The result is
\begin{eqnarray}
\Gamma(\Upsilon{\hspace{-0.5ex}}\to{\hspace{-0.5ex}}\tilde{S} \gamma)=
\hspace{-0.5ex}
\frac{256\pi^2e_b^2\alpha\alpha_s^2}{243}
\sin^2\theta_{\tilde{b}}\cos^2\theta_{\tilde{b}} 
\frac{
\langle \Upsilon| \mathcal{O}_1(^3S_1) |\Upsilon\rangle
\langle 0| \tilde{\mathcal{O}}_1^{\tilde{S}}(^1S_0) |0\rangle
     }{E_b^2 \tilde{E}_{\tilde{b}}^2 m_{\tilde{g} }^2}
(2M_{\tilde S})
\hspace{-0.5ex}
\left(1-\frac{M_{\tilde S}^2}{M_\Upsilon^2}\right).
\label{rate}
\end{eqnarray}

If one considers specific polarizations of the
$\Upsilon$, then the decay rate is no longer independent of the angle
$\theta$ between the photon and the axis that defines the direction of
longitudinal polarization. (In $\Upsilon$ production in $e^+e^-$
annihilation, for example, the polarization is transverse to the beam
direction.) In the case of equal population of the two transverse
polarization states, $d\Gamma/d(\cos\theta)=(3/8)(1+\cos^2\theta)\Gamma$,
while in the case of longitudinal polarization,
$d\Gamma/d(\cos\theta)=(3/4)(1-\cos^2\theta)\Gamma$, where $\Gamma$ is
found in Eq.~(\ref{rate}).

Using the nonrelativistic approximations $E_b\approx m_b$ 
and $\tilde{E}_{\tilde{b}}\approx m_{\tilde{b}}$
in Eq.~(\ref{rate}), 
we obtain the 
branching fraction 
\begin{eqnarray}
\textrm{Br}(\Upsilon\to\tilde{S} \gamma)
&=&
\frac{\Gamma(\Upsilon\to\tilde{S} \gamma)}
     {\Gamma(\Upsilon\to \mu^+ \mu^-)}\times\, 
\textrm{Br}(\Upsilon\to \mu^+ \mu^-)_{\textrm{Exp}}
\nonumber\\
&=&
\frac{64\pi\alpha_s^2}{81\alpha }
\sin^2\theta_{\tilde{b}}\cos^2\theta_{\tilde{b}}
\frac{
\langle 0| \tilde{\mathcal{O}}_1^{\tilde{S}}(^1S_0) |0\rangle
     }{m_b^2 m_{\tilde{b}}^2 m_{\tilde{g} }^2}\;M_\Upsilon^2\,M_{\tilde{S}}
\nonumber\\
&&\hbox{}\times
\left(1-\frac{ M_{\tilde S}^2 }{ M_\Upsilon^2  }\right)
\textrm{Br}(\Upsilon\to \mu^+ \mu^-)_{\textrm{Exp}}\,.
\label{bf}
\end{eqnarray}
In deriving Eq.~(\ref{bf}) we use
\begin{eqnarray}
\Gamma(\Upsilon\to \mu^+ \mu^-)=
\frac{8\pi e_b^2\alpha^2}{3}
\frac{\langle \Upsilon|\psi^\dagger\sigma^k\chi|0\rangle
\langle 0|\chi^\dagger\sigma^k\psi|\Upsilon\rangle
}{M_\Upsilon^2}
\end{eqnarray}
and take the vacuum-saturation approximation
\begin{equation}
\langle\Upsilon|{\cal O}_1({}^3S_1)|\Upsilon\rangle\approx
\langle \Upsilon|\psi^\dagger\sigma^k\chi|0\rangle
\langle 0|\chi^\dagger\sigma^k\psi|\Upsilon\rangle,
\end{equation}
which neglects terms of relative order $v^4$. Here, and throughout this 
paper, $v$ is the heavy-quark or heavy-squark velocity in the onium rest 
frame.
\begin{figure}
\begin{tabular}{cc}
\includegraphics[height=6cm]{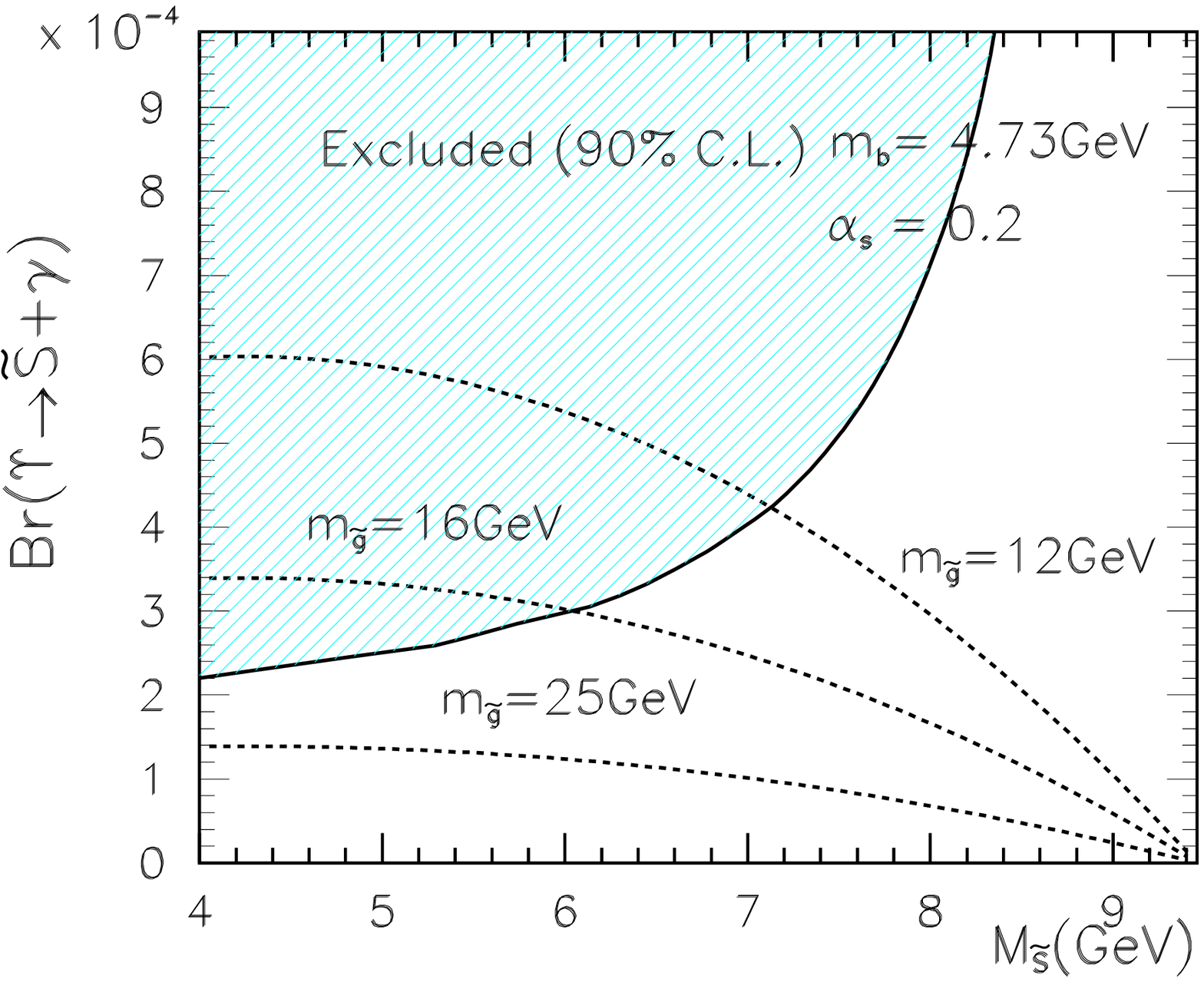}
\end{tabular}
\caption{
Branching fraction for the process 
$\Upsilon\to\tilde{S} \gamma$ as 
a function of $M_{\tilde{S}}$. 
The shaded area is excluded at the 90\% confidence level by the 
$\Upsilon\to X \gamma$ search of the CUSB Collaboration~\cite{CUSB-gamma}.}
\label{fig:BR}
\end{figure}

In the vacuum-saturation 
approximation, the color-singlet production matrix 
element $\langle 0| \tilde{\mathcal{O}}_1^{\tilde{S}}({}^1S_0) |0\rangle$
may be replaced by the corresponding decay matrix element
$\langle \tilde{S}| \tilde{\mathcal{O}}_1({}^1S_0) 
                 |\tilde{S}\rangle$, with uncertainties of relative order 
$v^4$. Furthermore, the sbottomonium decay matrix element $\langle \tilde{S}| 
\tilde{\mathcal{O}}_1({}^1S_0)|\tilde{S}\rangle$ is related to the 
heavy-quarkonium (HQ) matrix element of the same mass by 
\begin{eqnarray}
\langle \tilde{S}| \tilde{\mathcal{O}}_1({}^1S_0) 
|\tilde{S}\rangle \approx
\langle {\rm HQ}| \mathcal{O}_1({}^3S_1) |{\rm HQ}\rangle/2 ,
\label{wfn-sq}
\end{eqnarray}
where we neglect spin-dependent contributions to the HQ matrix element 
of relative order $v^2$.
The expressions on the left and right sides of Eq.~(\ref{wfn-sq}) are
proportional to the squares of the sbottomonium and quarkonium
wave functions at the origin, respectively. The value of the HQ matrix 
element is known at the $\Upsilon$ mass. We estimate its value at smaller 
quarkonium masses by assuming that it scales as $m_q^{3/2}$. This 
scaling behavior is approximately that which one obtains from the Martin 
potential \cite{martin}, which describes the $J/\psi$ and $\Upsilon$ 
systems reasonably well. Then we have
\begin{eqnarray}
\langle \tilde{S}| \tilde{\mathcal{O}}_1({}^1S_0) 
|\tilde{S}\rangle 
\approx
\left(\frac{m_{\tilde{b}}}{m_b}\right)^{3/2}
\langle \Upsilon| \mathcal{O}_1({}^3S_1) |\Upsilon\rangle/2.
\label{me-est}
\end{eqnarray}
We use a recent lattice measurement of the bottomonium matrix
element: 
$\langle \Upsilon| \mathcal{O}_1({}^3S_1) |\Upsilon\rangle%
=4.10\pm 0.42\hbox{ GeV}^3$ (Ref.~\cite{BKS-new}).

In Fig.~\ref{fig:BR}, we plot the branching fraction (\ref{bf}), for 
several values of $m_{\tilde g}$, as a 
function of $M_{\tilde S}\approx 2m_{\tilde b}$, 
using the estimate of the sbottomonium matrix element in Eq.~(\ref{me-est}). 
Here, and in all further numerical estimates, we take
$\sin^2\theta_{\tilde{b}}=1/6$, $m_b=4.73\pm 0.20$~GeV, $\alpha=1/137$,
$\alpha_s=0.2\pm 0.02$, and $\textrm{Br}(\Upsilon\to \mu^+
\mu^-)_{\textrm{Exp}}=2.48\pm 0.06\%$ (Ref.\cite{PDG}). This value 
for $\alpha_s$ corresponds approximately to the renormalization
scale $m_b$, which is an upper bound on the momentum transfer in the
radiative decay process.

In order to compare our result for the branching fraction with the
experimental resolution, it is necessary to know the width of the
$S$-wave sbottomonium state. The total width into light hadrons is
given, in leading order in $\alpha_s$, by the width into two gluons.
This quantity is computed, in leading order in the
nonrelativistic expansion, by Nappi \cite{nappi}:
\begin{equation}
\Gamma\left(\tilde{S}\to gg\right)
=
\frac{4\alpha^2_s}{3M_{\tilde{S}}^2}
 \left|R(0)\right|^2,
\label{2g}
\end{equation}
where $R(0)$ is the $\tilde{S}$ radial wave function at the origin.
Using Eq.~(\ref{me-est}) and taking $M_{\tilde{S}}\approx 2m_{\tilde{b}}$, we
have
\begin{eqnarray}
\left|R(0)\right|^2
&=&
\frac{4\pi}{N_c}
 \langle \tilde{S}|\mathcal{\tilde{O}}_1(^1S_0)|\tilde{S}\rangle.
\label{r02}
\end{eqnarray}
The width of the sbottomonium
state into light hadrons is less than 10~MeV in the range of
parameters proposed in the light-bottom-squark scenario~\cite{Berger:2000mp}. 
This width is
less than the energy resolution in the CUSB search for monochromatic
photon signals \cite{CUSB-gamma}. Therefore, we can compare our estimate
of the branching fraction $\textrm{Br}(\Upsilon\to\tilde{S} \gamma)$
directly with the CUSB 90\% confidence level for the exclusion of
$\Upsilon\to X \gamma$, which is plotted, along with our
estimate, in Fig.~\ref{fig:BR}.  
We see that, if a bound state is formed,
then a part of the range of mass parameters proposed in the 
light-bottom-squark scenario is disfavored.

There are several uncertainties in our calculation. The uncertainty in
$\hbox{Br}(\Upsilon\rightarrow\mu^+\mu^-)$ is 2.4\%. The uncertainty in
$m_b$ is 4.2\%. The uncertainties in the value of $\alpha_s$ and in the
lattice computation of the bottomonium matrix element are each about
10\%. There is also an uncertainty from the extrapolation of the HQ matrix
element from $M_\Upsilon$ to $M_{\tilde S}$. We estimate it by checking
the accuracy of the extrapolation against the phenomenological values of
the  wave functions at the origin for the $\Upsilon$ and the $J/\psi$,
as determined from the data for the decay rates into lepton pairs
combined with the next-to-leading-order QCD expressions for those decay
rates. We conclude that the extrapolation error is approximately
$31\%\times (M_\Upsilon-M_{\tilde S})/(M_\Upsilon-M_{J/\psi})$. There are
uncalculated relativistic corrections of the order of the square of the
heavy-quark or squark velocity in the onium rest frame. We estimate
these to yield an uncertainty in the decay rate of $20\% \times
(M_\Upsilon-M_{\tilde S})/(M_\Upsilon-M_{J/\psi})+10\%$. There are
uncertainties from uncalculated corrections of higher order in
$\alpha_s$ and from the imprecision in the value of the renormalization
scale. We estimate these by varying the scale of $\alpha_s$ from $m_b/2$
to $2m_b$. This procedure yields uncertainties in $\alpha_s$ of $+33\%$ and
$-18\%$. As $M_{\tilde S}$ approaches $M_\Upsilon$, the momentum
transfer in the radiative decay becomes considerably less than $m_b$,
and, in this region, our choice of scale probably results in an
underestimate of the decay rate. In this same region, we expect to find
violations of the NRQCD factorization, which holds only for
$M_\Upsilon-M_{\tilde S}\gg \Lambda_{\rm QCD}$. 

\begin{figure}
\includegraphics[height=6cm]{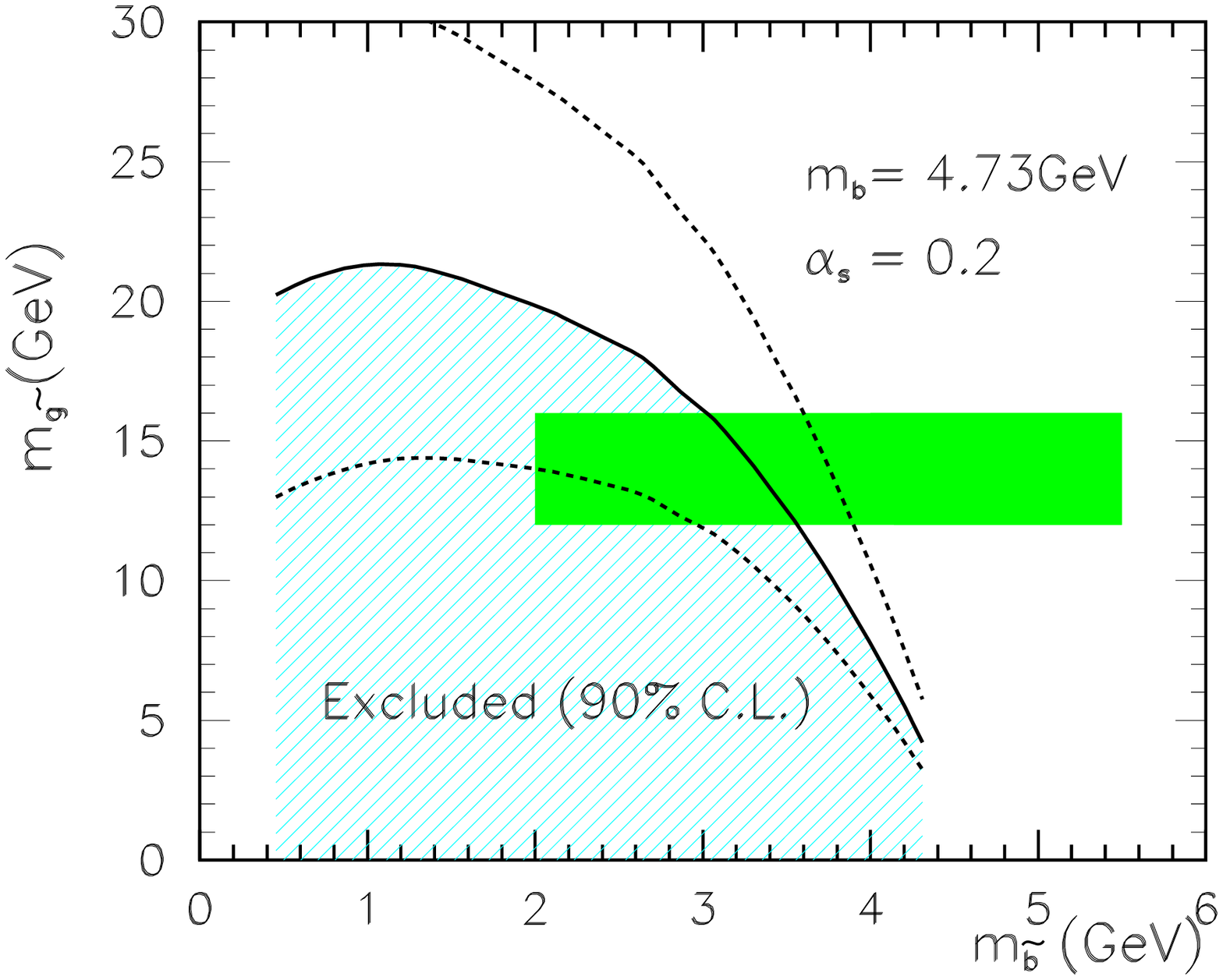}
\caption{
The regions of the $m_{\tilde{b}}$-$m_{\tilde{g}}$ parameter space that
are excluded at the 90\% confidence level by the $\Upsilon\to X \gamma$
search of the CUSB Collaboration (shaded region)~\cite{CUSB-gamma}.
The solid curve represents
the central value of the theoretical calculation, and the dashed curves
show the uncertainties on the theoretical values, as described in the text.
The strip shows the region $2<m_{\tilde{b}}<5.5$ GeV, $12<m_{\tilde{g}}<16$ GeV
proposed in the light-bottom-squark scenario~\cite{Berger:2000mp}.}
\label{fig:EXCLUDE}
\end{figure}
In Fig.~\ref{fig:EXCLUDE} we show the region excluded by the CUSB data.
We calculate the uncertainty band on the boundary of the excluded region
by adding, in quadrature, the theoretical uncertainties mentioned above.
 We also plot the values of $m_{\tilde g}$ and $m_{\tilde b}$ that are
suggested in the light-bottom-squark scenario~\cite{Berger:2000mp}. At
$m_{\tilde g} = 12$ GeV, provided that the bottom squark lifetime is
great enough to permit formation of the $\tilde{b} \tilde{b}^*$ bound
state, the mass range $m_{\tilde b} < 3.5^{+0.4}_{-0.6}\hbox{~GeV}$ is
excluded at the 90\% confidence level by the CUSB data. At $m_{\tilde g} =
16$ GeV, the central and upper values of the excluded range are
$m_{\tilde b} = 3.0$ and $3.6$~GeV, but the theoretical uncertainties do
not permit us to specify a lower limit at the 90\% confidence level. One can
probe the region of higher bottom-squark masses by increasing the
statistics of the photon sample and by examining decays from bottomonium
states of higher mass, such as the $\Upsilon(3S)$ and the $\Upsilon(4S)$.
\footnote{In order to apply the expression for the
branching fraction in Eq.~(\ref{bf}) to decay from a state other than
the $\Upsilon$, one must replace $\textrm{Br}(\Upsilon\to \mu^+
\mu^-)_{\textrm{Exp}}$ with the rate for the decaying state and replace
$m_b$ with one-half the mass of the decaying state.}

Bottom squarks with mass $2~\hbox{GeV} < m_{\tilde{b}} <5.5~\hbox{GeV}$
along with gluinos with mass $12~\hbox{GeV} < m_{\tilde{g}}
<16~\hbox{GeV}$ are proposed in Ref.~\cite{Berger:2000mp} to explain the
larger-than-predicted rate for bottom quark ($b\bar{b}$) production at
the Fermilab collider.  In this Letter, we show that CUSB data on
radiative $\Upsilon$ decays already provide an important additional
constraint on the mass ranges. The high-statistics 2002 CLEO data should
either permit discovery of squarkonium bound states, confirm the
exclusion region of the earlier CUSB data, or further narrow the allowed
range of supersymmetry parameter space.

\begin{acknowledgments}
We acknowledge valuable conversations with D.~Besson, R.~Galik, Jik Lee, 
S.~Nam, D.~Son, and M.~Tuts.  
Work in the High Energy Physics Division at Argonne National Laboratory 
is supported by
the U.~S.~Department of Energy, Division of High Energy Physics, under
Contract No.~W-31-109-ENG-38. 
\end{acknowledgments}

\end{document}